%%%%%%%%%%%%%%%%%%%%%%%%%%%%%%%%%%%%%%%%%%%%%%%
\input harvmac.tex
%\draft
\Title{\vbox{\baselineskip12pt \hbox{IFT-UAM/CSIC-98-7}
}}
{\vbox{\centerline{On the $SL(2,Z)$ covariant}
\bigskip
\centerline{
World-Sheet Action with 
Sources}}} \centerline{\bf Sudipta Mukherji}
\smallskip\centerline{\it Instituto de F\'isica Te\'orica, C XVI,}
\smallskip\centerline{\it Universidad Aut\'onoma de Madrid, Madrid 28049,
Spain}
\smallskip\centerline{e-mail: mukherji@delta.ft.uam.es}
\vskip .3in
We analyse various world-sheet properties of the $SL(2,Z)$ covariant type
IIB string action by coupling it with $SL(2,Z)$ covariant source.

%%%%%%%%
\Date{11/98}
%%%%%
\def \cf {{\cal F}}
\def \cm {{\cal M}}
\def \ctf {\tilde{\cal F}}

%%%%%%%%
\lref\gkp{S. Gukov, I. Klebanov and A. Polyakov, {\it Dynamics of (n,1)
strings}, hep-th/9711112.}

\lref\ck{C. Callan and I. Klebanov, {\it D-brane boundary dynamics},
hep-th/9511173.}

\lref\witten{E. Witten, {\it Bound states of strings and $p$-branes}, 
hep-th/9510135.}

\lref\schwarz{J. Schwarz, {\it An $SL(2,Z)$ multiplet of type IIB 
superstrings}, hep-th/9508143.}

\lref\schwarzt{J. Schwarz, {\it Lectures on superstring and M theory 
dualities}, hep-th/9607201.}

\lref\schmi{C. Schmidhuber, {\it D-brane actions}, hep-th/9601003.}

\lref\zeid{M. Abou Zeid and C. Hull, {\it Intrinsic Geometry of 
D-branes}, hep-th/9704021.}

\lref\bergman{O. Bergman, {\it Three-pronged strings and 1/4 BPS states 
in N = 4 Super-Yang-Mills Theory}, hep-th/9712211.}

\lref\deser{S. Deser, A.Gomberoff, M. Henneaux and C. Teitelboim, {\it 
p-brane dyons and electric-magnetic duality}, hep-th/9712189.}

\lref\wester{M. Cederwall and A. Westerberg, {\it World-volume fields, 
$SL(2,Z)$ and duality: The type IIB three-brane}, hep-th/9710007.}

\lref\lozano{Y. Lozano, 
{\it D-brane dualities as canonical transformations}, hep-th/9701186.}

\lref\town{ P.K. Townsend, {\it Membrane tension and manifest $IIB$
$S$-duality}, 
hep-th/9705160.}

\lref\cede{M. Cederwall and P.K. Townsend, {\it The manifestly
$SL(2,Z)$-covariant superstring}, hep-th/9709002.}

%\lref\deser{S. Deser, A. Gomberoff, M. Henneaux and C. Teitelboim,
%{\it Duality, self-duality, sources and charge quantization in
%abelian $N$-form theories}, hep-th/9702184.}

\lref\berman{D. Berman, {\it $SL(2,Z)$ duality of Born-Infeld 
theory from non-linear self-dual electrodynamics in six dimensions},
hep-th/9706208.}

\lref\khoudeir{A. Khoudeir and Y. Parra, {\it On duality in the 
Born-Infeld theory}, hep-th/9708011.}

\lref\cederwall{M. Cederwall and A. Westerberg, {\it World-volume fields, 
$SL(2,Z)$ and duality: the type IIB 3-brane}, hep-th/9710007.}

\lref\dasgupta{K. Dasgupta and S. Mukhi, {\it BPS nature of 3-string 
junctions}, hep-th/9711094.}

\lref\sen{A. Sen, {\it String network}, hep-th/9711130.}

\lref\zwiebach{M. Gaberdiel and B. Zwiebach, {\it Exceptional groups from 
open strings }, hep-th/9709013.}

\lref\tseytlin{A.A. Tseytlin, {\it Self-duality of Born-Infeld action and 
Dirichlet 3-brane of type IIB superstring theory}, hep-th/9602064.}

%%%%%%%%%%%%%%%%%%%%%

\noindent{\bf 1.}~Ten dimensional type IIB string theory is conjectured to 
have $SL(2,Z)$
symmetry. Under generic $SL(2,Z)$ transformations, the
NS-NS and the R-R charges of the theory mix among each other. As a 
result, in  particular, a macroscopic fundamental string (F-string),
under $SL(2,Z)$, gets transformed to a D-string or various bound states 
of F and D-strings~\refs{\schwarz, \witten } . Various such objects can be 
classified by identifying their tensions as 
\eqn\tension{ T_{(p,q)} = {\sqrt{ (p - q\chi )^2 + {q^2\over g^2}}},}
where $p,q$ are relatively prime integers, $g$ is the type IIB string
coupling related to the expectation value of the dilaton and $\chi$ is
the R-R scalar. The F-string is identified with tension $T_{(1,0)}$ and 
D-string with $T_{(0,1)}$. The rest can, in turn, be thought of 
as the bound states of these two and, following the literature, will
be called $(p,q)$ strings. It is known that, for a macroscopic 
F or D-strings solutions, the world sheet F and D-string actions act as
sources. However, since F and D-string solutions are related by $SL(2,Z)$ 
transformation, one would expect to see a version of this transformation 
relating world-sheet actions of F and D-strings. This is indeed the case 
as has been discussed in \refs{\schmi, \tseytlin, \lozano, \zeid }. 
Moreover, in subsequent development, a 
manifest covariant action has been proposed which describes the entire 
orbit of $(p,q)$ string \refs{\town, \cede }. Here, in the rest of the 
note, we will use  this 
action in order to explore some aspects of world-sheet physics of the 
$(p,q)$ strings. 

It is, by now, well known that \refs{\schwarzt, \zwiebach , \dasgupta }, if 
certain rules are satisfied, the type 
IIB string theory in ten dimension admits stable configurations of three or
more string junctions. Furthermore, combining these junctions in 
a suitable way, it is possible to construct stable string network 
\refs{\sen}. 
Whenever a string ends on another, the end point of one acts as a 
source for the other on its world-sheet. This naturally leads us to 
the investigation of string world-sheet action in presence of sources. In 
particular, we would analyse the $SL(2,Z)$ covariant action in presence 
of such sources and discuss the consequences.

\bigskip

\noindent{\bf 2.}~As mentioned above, the $SL(2,Z)$ covariant IIB string 
action has been recently proposed in \refs{\town} and the
$\kappa$-symmetry of the corresponding
supergravity backgrounds has been analyzed subsequently
in \refs{\cede}. The action is constructed by introducing
two 1-form gauge potentials, $A_\mu$ and $\tilde A_\mu$.
$A_\mu$ is the usual Born-Infeld (BI) field that appear on the
standard D-brane worl-sheet action. On the other hand, 
$\tilde A_\mu$ plays the role of BI field on the fundamental 
type IIB string. This field, however, does not appear
in standard wold-sheet action of fundamental string
but, as we will see, $S$-duality covariance, in various ways, requires
such a field. Specific choices of $(A_\mu , \tilde A_\mu)$ lead
to D- or F-strings. This, in turn, breaks $SL(2,Z)$ covariance
on the world-sheet.

The duality covariant action is then straight forward to construct.
As usual, first one defines modified two-form fields
\eqn\twoforms{\eqalign{&\cf = \epsilon^{\mu \nu} \cf_{\mu \nu}
              = \epsilon^{\mu \nu} (\partial_\mu A_\nu - B_{\mu \nu})\cr
              &\ctf = \epsilon^{\mu \nu} \ctf_{\mu \nu}
              = \epsilon^{\mu \nu} (\partial_\mu {\tilde A}_\nu - 
              {\tilde B}_{\mu \nu}),}}
with  $B_{\mu \nu}$ and ${\tilde B}_{\mu \nu}$ being the pullbacks
to the world-sheet of the NS-NS and R-R two form gauge potentials
respectively. With these set of fields, the $SL(2,Z)$ covariant
IIB action is
\eqn\covar{ S = \int d\tau d\sigma{1\over{2v}}[ {\rm det} g
          + e^{-\phi} \cf^2 + e^{\phi}(\ctf -\chi \cf)^2].}
Here $g$ is the induced metric in the Einstein frame. The 
string frame metric $g_s$ is related to $g$ as $g_s = e^{\phi\over 2}g$.
Hence \covar\ in string frame will have an extra factor of $e^{-\phi}$
infront of ${\rm det} g_s$. 
Now, if we define 
\eqn\matr{\tau = \chi + i e^{-\phi},~~~~{\bf F} = \pmatrix{\cf\cr \ctf},~~~~
\Lambda = \pmatrix{p & r\cr q & s}, ~{\rm with}~
ps -qr = 1,}
we see that the action is invariant if the fields transform as
\eqn\trans{ \tau^\prime = {p\tau + r\over{q\tau + s}},~~~~
{\bf F}^\prime = \Lambda {\bf F}.}
In \matr\ , $\chi$ is the usual R-R scalar, $\phi$ is the dilaton
and  $\Lambda$ 
is an $SL(2,Z)$ matrix. Moreover, we note that in \covar\ , $v$ is 
an auxiliary field and, as we will see below, its  expectation value 
determines the tension of the string.

The action \covar\ can be written in a manifestly covariant
way by introducing a matrix $\cm$
\eqn\mmatrix{\cm = e^{\phi}\pmatrix{\tau^2 & -\chi \cr 
                            -\chi & 1}}
as 
\eqn\manifest { S = \int d^2\sigma{1\over{2v}}[ {\rm det} g
                    + {\bf F}^{\rm T}\cm {\bf F}].}
Using the first equation of \trans\ , it is easy to check that
under $SL(2,Z)$, $\cal M$ transforms as ${\cal M} \rightarrow 
(\Lambda^T)^{-1}{\cal M} \Lambda^{-1}$. Thus the second term
in \manifest\ is indeed $SL(2,Z)$ invariant.
The equation of motion for $\tilde A$ that follows from \covar\
is
\eqn\vtilde {\ctf -\chi\cf = e^{-\phi} v T,}
where $T$ is a constant. If we now substitute \vtilde\ in \covar\
and add a total derivative term
\eqn\deriv{-\int T d\tilde A = -T\int (\ctf + \tilde B ),}
we get
\eqn\nextstep{S = -\int [{e^{-\phi}~{\rm det}g_s\over {2v}} + 
{e^{-\phi}\cf^2\over {2v}} - {e^{-\phi}vT^2\over 2}] - T\int \chi \cf
- T \int \tilde B.}
In writing down the last equation, we have used \vtilde .
Now solving for $v$ and substituting it back to \nextstep\ , we get 
\eqn\dbrane{ S = - T \int e^{-\phi}{\sqrt{- {\rm det}( g_s + \cf)}} - T
\int ( \chi \cf + \tilde B ),}
which is the standard D-string action.

It is also easy to get a $(p,q)$ string for the above action.
For that, one needs to simply integrate over $A$ and $\tilde A$.
A detail discussion on this can be found in the original papers
\refs{\town,\cede}.

Before we go further, let us note that we can fix a static gauge.
The action \manifest\ in the string metric takes the following
form:
\eqn\staticg{ S = \int d^2\sigma{1\over{2v}}[ e^{-\phi}{\rm det} 
( \eta_{\mu\nu} + \partial_\mu\phi^i \partial_\nu\phi^i )
                    + {\bf F}^{\rm T}\cm {\bf F}].}
Here, in writing down the action, we have identified the space 
time co-ordinates $X^0$ and $X^1$ with $\tau$ and $\sigma$ respectively.
On the other hand, $\phi^i$ corresponds to the fluctuation of eight transverse
directions to the world-sheet. Now we notice that any constant scaling
of the world sheet metric can be absorbed by simply rotating the 
world-sheet by a constant angle with respect to the target space 
co-ordinates $X^0$ and $X^1$. Since field strengths corresponding to
gauge fields are constants in two dimension, such overall scaling 
can be obtained simply by turning on the gauge fields. In what follows,
we analyse this in detail by exploiting certain properties of the covariant 
action 
in the presence of $SL(2,Z)$ covariant source term.  As we will see, 
among others, the effect of the source term is to induce  variations in 
${\cal F}_{\mu\nu}$ and/or ${\ctf}_{\mu\nu}$. This, in turn, 
changes the shape of the world-sheet. We hope to bring out the advantages 
of working in a $SL(2,Z)$ covariant frame work as we go along.

\bigskip

\noindent{\bf 3.}~~Since \manifest\ contains charged fields, it is
natural to study the action in presence of sources. However, as mentioned
before, such sources appear naturally if a F string (D-string)
ends on a D-string (F-string). 
With the simplest covariant choice of such a source term, the action 
takes the following form:
\eqn\complete{ S = \int d\sigma d\tau{1\over{2v}}[ {\rm det} g  
                    + {\bf F}^{\rm T}\cm {\bf F}] + \int d\tau{\bf 
                    J}^{\rm T}{\bf A_\tau}.}
Here the matrix $\bf J$ has the form
\eqn\scource{{\bf J} = \pmatrix {p\cr q}.}
We also note that under $SL(2,Z)$, $\bf J$ transforms as 
$J^{\rm T} \rightarrow J^{\rm T}\Lambda^{-1}.$
The last term in \complete\ acts as a source term to the original
action. This term can in turn be written as an integral over 
world sheet with a delta function support along $\sigma$. 

In order to study the effect of the source term, we start with the
simplest case.  We  set the R-R scalar $\chi$,
the R-R and NS-NS two forms to zero.
The action for a D-string in this case is given by \manifest\
with 
\eqn\ds{\cf = 0, ~~~~ \ctf = v e^{-\phi}.}
We suppose that a D-string interacts with a source term corresponding to
${\bf J} = \pmatrix{1\cr 0}$.
Now, the equations of motion for $A$ and $\tilde A$ that follows from
\complete\ have the forms
\eqn\deq{\cf = -e^{\phi} v, ~~~~\ctf = v e^{-\phi}.}
It is now easy to check that if we substitute \deq\ in \manifest\
and integrate over $v$, we get an action for (1,1) string with tension
$T_{(1,1)} = {\sqrt {1 + e^{-2\phi}}}$. 

One benefit of working with
a covariant action is immediate. We start with a (1,0) string and 
${\bf J} = \pmatrix{0\cr 1}$ as source instead. It is easy to check that  
(1,0) string, given by $\cf = v e^{-\phi} , ~ \ctf = 0$ will again
turn into a (1,1) string with $\cf$ and $\ctf$ given by \deq. Notice
that the above two examples are related to each other by a $SL(2,Z)$
metrix. The above result generalizes trivially to
 arbitrary  integer $(p,q)$.

Similar story repeats when one includes the R-R scalar $\chi$. We study 
one example here. For a (0,1) string, 
\eqn\axd{\cf = v\chi e^\phi, ~~~~\ctf = ve^{-\phi} + v \chi^2e^\phi.}
In the presence of a source with ${\bf J} = \pmatrix{1\cr 0}$, it again
turns in to a (1,1) string as before but with modified fields
\eqn\axo{\cf = v(\chi - 1) e^\phi, 
~~~~\ctf = ve^{-\phi} + v \chi (\chi- 1)e^\phi.} as can be seen from
\complete. Substituting thus in \manifest\ , we get a string action 
with tension ${\sqrt {(1 - \chi )^2 + e^{-2\phi}}}$ as expected.

Next, we consider various field configurations which describe
strings around the source. 
Again for simplicity, we consider a $(0,1)$
string turning into a $(1,1)$ string. 

We take the $(0,1)$ string with non-vanishing $\phi$ and $\chi$
oriented along the $y$ axis of the two dimensional $x-y$ plane.
The vector fields that describe this string are easy to find.
\eqn\config{{\bf F}^{(0,1)} = \pmatrix{v\chi_0 e^\phi_0\cr v(e^{-\phi_0}+ 
\chi_0^2e^{\phi_0})} = \pmatrix{\chi_0\over{\sqrt{\chi_0^2 + e^{-2\phi_0}}}
\cr {\sqrt{\chi_0^2 + e^{-2\phi_0}}}}.} 
Here, in the second expression, we have substituted the value
of $v$ that follows from its equation of motion. Moreover, the subscript
on $\phi$ and $\chi$ denotes their constant background expectation values.

Since we have taken the D-string to be oriented along the $y$ axis,
it is easy to get the gauge fields by integrating \config\ along the axis.
\eqn\configv{{\bf A}^{(0,1)}_0 = \pmatrix{A^{(0,1)}_0\cr \tilde 
A^{(0,1)}_0} = \pmatrix{y\chi_0\over{\sqrt{\chi_0^2 + e^{-2\phi_0}}}
\cr y{\sqrt{\chi_0^2 + e^{-2\phi_0}}}}.}
Similarly, the $(1,1)$, that is produced because of the presence of the
source, have the following field configuration (follows from \axo\ )
\eqn\configpq{{\bf A}^{(1,1)}_0 = \pmatrix{A^{(1,1)}_0\cr \tilde 
A^{(1,1)}_0} = \pmatrix{{z (\chi_0 - 1 )\over{\sqrt{(1-\chi_0 )^2 + 
e^{-2\phi_0}}}}\cr {z[e^{-2\phi_0} + (\chi_0 -1 )\chi_0 ]\over{\sqrt{(1 
-\chi_0)^2 + e^{-2\phi_0}}}}    }.}
Here we have chosen $z$ to be the direction of the $(1,1)$ string in the
$x-y$ plane. It is now clear that since there is no source for
$\tilde A$, $\tilde A^{(1,1)}_0 =  \tilde A^{(0,1)}_0$.
%(This follows
%from integrating the corresponding electric field $\tilde E$ around the 
%source).
Comparing now with \configv\ and \configpq, we see that the $(1,1)$ 
string is 
bent making an angle $\theta_1$ with the original D-string. The angle
is given by 
\eqn\angleo{{\rm cos}\theta_1 = 
{e^{-2\phi_0} + (\chi_0 - 1 )\chi_0 \over {\sqrt{[(1- \chi_0 )^2 + 
e^{-2\phi_0}][\chi_0^2 + e^{-2\phi_0}]}}}.}
However, we see that the string tension around the source
is only balanced (required for the stability of the configuration)
if we consider the following simple possibility. We take the
source to be the end point of a F-string extending in the same
two dimensional plane making an angle $\theta_2$ with the D-string 
such that
\eqn\anglet{{\rm cos}\theta_2 = {\chi_0\over {\sqrt{\chi_0^2 + 
e^{-2\phi_0}}}}.}
This is thus the three string junction of \refs{\dasgupta}. However,
we would like to stress here that we found it out by studying
the dual gauge field $\tilde A$, which is a simple consequence of working 
with a  $SL(2,Z)$ covariant action. Moreover, since we know 
the transformation properties of  $\bf F$ and $\bf J$
under $SL(2,Z)$, we can obtain a generic junction configuration simply by
applying $SL(2,Z)$ matrix on \configv\ and the source term.

In a similar manner, we can study the effect of ${\bf J} = 
\pmatrix{n\cr 0}$ source on the D-string. The resultant $(n,1)$ string 
makes an angle $\theta_1$ with the D-string given by
\eqn\anglen{{\rm cos}\theta_1 =
{e^{-2\phi_0} + (\chi_0 - n )\chi_0 \over {\sqrt{[(n- \chi_0 )^2 +  
e^{-2\phi_0}][\chi_0^2 + e^{-2\phi_0}]}}},}
instead of \angleo. Furthermore, if we consider the source to be
the end point of a $(n,0)$ string, the tension around the junction
is only balanced when $(n,0)$ string makes an angle $\theta_2$ given by
\anglet\ with the D-string. It is interesting to note that in the 
large $n$ limit, various angles around the junction are independent
of $n$ and determined completely by the background values of $\chi$ and 
$\phi$. The angle is given by 
\eqn\anglett{{\rm cos}\theta_1 = {\chi_0\over {\sqrt{\chi_0^2 +
e^{-2\phi_0}}}}.}
It is precisely this limit when the world-sheet electric 
field reaches its critical
value \refs{\gkp}. If we further add open strings to the $(n,1)$ string, the 
angle is hardly changed. This indicates that in the large $n$ limit, the
open string that ends on the $(n,1)$ string becomes tensionless. From
\anglett\ it also follows that if the 
R-R field is set to zero, the $(n,1)$ string
is completely bent making an angle $\pi\over 2$ with the original D-string 
and is independent of the string coupling $e^{\phi_0}$. 

From the above discussion, therefore, question naturally arises as to what 
exactly happens to the open string that end on 
$(n,1)$ string as we increase $n$? 
Using the boundary state formalism \refs{\ck}, it is possible to identify 
certain set of operators in the correspoding conformal field theory which
generates deformations along the direction of the $(n,1)$ string. 
It turns out that, in the large $n$ limit, these deformations correspond to 
pure gauge deformations and hence decouple from the theory. 
A detail discussion of
this issue will be presented elsewhere.

Finally, we would like to make the following comments. In type IIB string
theory, besides strings, we have other branes as solitons. Among them,
the three brane is self-dual under $SL(2,Z)$. At the level of 
world-volume theory, it has been analyzed in \refs{\tseytlin}. 
Furthermore, in \refs{\deser}, a detail analysis of self-duality properties 
was carried out for four dimensional Maxwell electrodynamics. 
Generalizing their result for the three-brane world volume theory is 
straight forward. A preliminary analysis shows that external 
electro-magnetic sources can be introduced on the world-volume 
maintaining self-duality properties of the three-brane. In the context of 
type IIB string theory, it has the following interpretation. The end 
point of a $(p,q)$ string acts as source of electro-magnetic field when 
it ends on a three-brane. Thus it is possible to have string and 
three-brane junction preserving the self-dual properties of the 
three-brane (for a detail discussion of such junction configurations, see 
for example \refs{\bergman}). Recently, in \refs{\wester} , a manifestly 
$SL(2,Z)$ 
invariant three-brane world-volume action has been proposed by again 
introducing auxiliary gauge field. It will thus be of interest to 
understand the world-volume physics by coupling it with $SL(2,Z)$ 
invariant source. We hope to report on it in the future.

\noindent {\bf Acknowledgements}: I would like to thank  C\'esar 
G\'omez, Patrick Meessen and especially Tom\'as Ort\'in for useful 
discussions. The work is supported by Ministerio de Educaci\'on y
Cultura of Spain and also through the grant CICYT-AEN 97-1678.
\vfill\eject
\listrefs
\bye